# Space Charge Limited Current with Self-heating in $Pr_{0.7}Ca_{0.3}MnO_3$ based RRAM

I. Chakraborty[1], N. Panwar[1], A. Khanna[1] and U. Ganguly[1]

[1]*Department of Electrical Engineering, Indian Institute of Technology Bombay, Mumbai, 400076, India*

**Abstract:** Space Charge Limited Current (SCLC) based conduction has been identified for PCMO-based RRAM devices based on the observation that $I \propto V^\alpha$ where $\alpha \approx 2$. A critical feature of the *IV* characteristics is a sharp rise in current ($\alpha \gg 2$) which has been widely attributed to trap-filled limit (TFL) followed by an apparent trap-free SCLC conduction. In this paper, we show by TCAD analysis that trap-filled limit (TFL) is insufficient to explain the sharp current rise ($\alpha \gg 2$). As an alternative, we propose a shallow trap SCLC model with self-heating effect based thermal runaway to explain the sharp current rise followed by a series resistance dominated regime. Experimental results over a range of 25°C-125°C demonstrate all 4 regimes (i) Ohmic ($\alpha = 1$), (ii) shallow trap SCLC ($\alpha \approx 2$), (iii) current shoot up ($\alpha \gg 2$) and (iv) series resistance ($\alpha = 1$). Further, TCAD simulations with thermal modeling are able to match the experimental *IV* characteristics in all the regimes. Thus, a current conduction mechanism in PCMO-based RRAM supported by detailed TCAD model is presented. Such a model is essential for further quantitative understanding and design for PCMO-based RRAM.

Manganite like $Pr_{0.7}Ca_{0.3}MnO_3$(PCMO) based bipolar RRAM is attractive for non-stochastic switching[1], area-dependent current scaling[2-3], fast switching, excellent endurance, and forming-less operation[4-5]. Among various mechanisms suggested to explain the *IV* characteristics for PCMO-based RRAM including Schottky barrier-like interface resistance switching[6], and Metal-Insulator Mott-Transition[7-8], SCLC-based mechanism has been a strong contender[9-14]. We have recently developed a quantitative analysis methodology for the extraction of uniform trap density ($N_T$) and single trap energy ($E_T$) for PCMO RRAM[15] based on SCLC theory[16]. The *IV* characteristics may be described by $I \propto V^\alpha$ where an initial Ohmic region ($\alpha = 1$) is followed by an SCLC region ($\alpha \approx 2$). Then, a critical feature of the *IV* characteristics is a sharp rise in current ($\alpha \gg 2$) which has been attributed largely to trap-filled limit (TFL) followed by an apparent trap-free SCLC conduction[17-21] based on "curve-fitting" analysis. In fact, a sharp (<20mV/decade) current rise of more than 2 orders has been observed in the Low Resistance State (LRS) that enables selector-less PCMO RRAM[22-23]. However, such an SCLC model indicated by "curve fitting" needs to be verified by detailed physical modeling. In this paper, we analyze the *IV* characteristics of PCMO-based RRAM using TCAD modeling to demonstrate that TFL is insufficient to explain the sharp current rise ($\alpha \gg 2$). We propose a single-trap SCLC-based model with self-heating followed by series resistance limited current to be sufficient to explain the observations. As validation, experimental *IV* characteristics from 25°C-125°C are compared to simulated *IV* characteristics to demonstrate that essential features are quantitatively reproduced.

A 64nm layer of $Pr_{0.7}Ca_{0.3}MnO_3$(PCMO) was deposited by Pulsed Laser Deposition (PLD) at room temperature on a substrate consisting of bottom electrode of Pt (57nm)/Ti (9nm) deposited on $SiO_2$/Si substrate[1,4]. The device was annealed at 650°C in $N_2$ for 120s. Tungsten probe-tips were used as top electrode for facile characterization[1, 4, 24]. Agilent B1500 was used to measure *IV* characteristics. Typical *IV* characteristics, shown in Fig. 1(a) at various current compliances ($I_{COMPLIANCE}$) indicate that the current level of LRS increases with $I_{COMPLIANCE}$ and saturates.

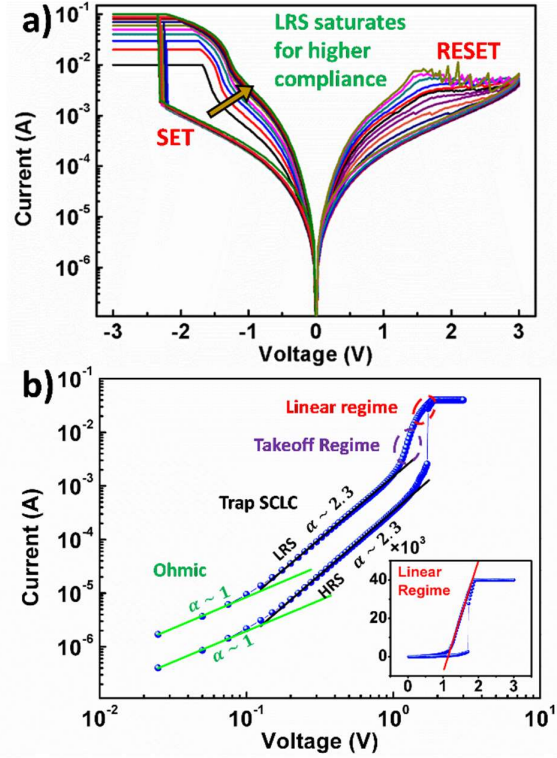

Fig. 1. (a) *IV* characteristics for different $I_{COMPLIANCE}$. LRS resistance decreases with increase in $I_{COMPLIANCE}$. (b) log-log plot of *IV* characteristic exhibiting negative LRS and HRS comprising of different regimes of current transport. Inset shows linear plot of *IV* characteristic with fit of the linear regime (marked as red).

Fig. 1(a) shows the *IV* characteristics of our device at different $I_{COMPLIANCE}$ exhibiting two resistance states LRS and High Resistance State (HRS). The log-log plot in Fig. 1(b) of *IV* characteristic in negative LRS and HRS shows 4 distinct regimes based on the exponent α of V i.e. $I \propto V^{\alpha}$: (i) Ohmic (α = 1) and (ii) SCLC (α ≈ 2) at low biases which indicates shallow trap SCLC. This is followed by (iii) sharp slope (α ≫ 2) and (iv) a saturated slope (α ≤ 2) followed by (v) current compliance (I = constant). As regime (iii) is normally attributed to TFL[17-21], we explore the behavior using TCAD simulations.

To qualitatively compare experimental *IV* characteristics with trap-SCLC *IV* characteristics, we performed simulations of a M-I-M structure using Sentaurus™ TCAD to demonstrate the qualitative difference between various types of SCLC. A band diagram of p-type PCMO with shallow and deep traps is depicted in Fig. 2(a) at (i) equilibrium and (ii) under bias. Trap-free SCLC-based *IV* characteristics exhibits a typical Ohmic (α = 1) followed by SCLC (α ≈ 2) in Fig. 2(b). The presence of traps modifies the *IV* characteristics. For *shallow trap SCLC*, regime (ii) i.e. α ≈ 2 is prominently preserved. Based on SCLC theory, both trap occupation and valence band occupation is given by Boltzmann distribution. The ratio of charge in valence band ($n$) vs. shallow trap ($n_{TS}$) is given by

$$\theta = \frac{n}{n_{TS}} = \frac{N_V \exp\left(\frac{E_V - E_F}{kT}\right)}{N_T \exp\left(\frac{E_{T-S} - E_F}{kT}\right)} = N_V/N_T \exp\left(\frac{E_V - E_{T-S}}{kT}\right) \quad \dots (1)$$

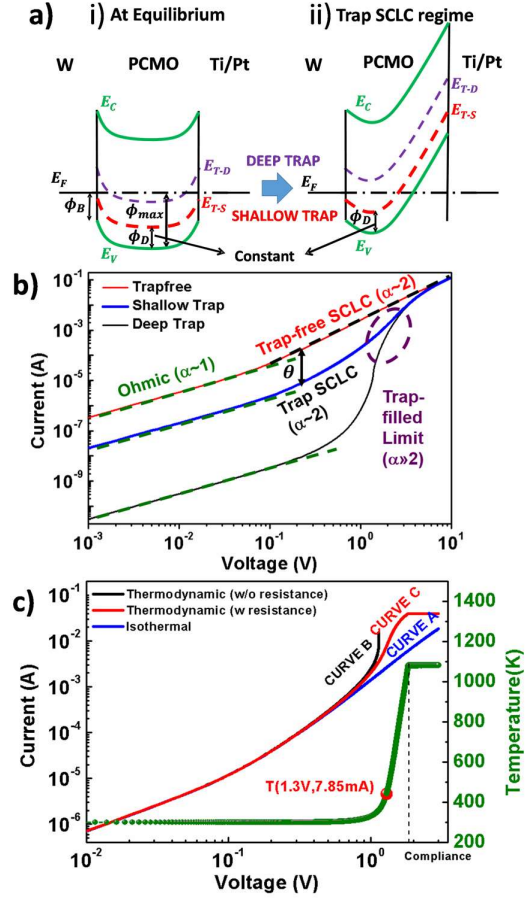

Fig 2: (a) Band profile comparison between (i) Ohmic regime and (ii) Trap-SCLC regime showing deep vs. shallow trap energy. Shallow traps can never be completely filled at high bias as they will partially lie above "injecting" Fermi Level (b) Simulated $IV$ characteristics in presence of shallow, deep and no traps (c) Simulated $IV$ showing the difference between isothermal simulations (blue) and thermodynamic simulations with thermal runaway (black), and thermal runaway arrested by the effect of series resistance (red). The maximum internal device temperature i.e. $T_{max}$ (green) increases steeply beyond a critical power just as the sharp ($\alpha \gg 2$) $IV$ characteristics is observed.

Here, as $\theta \ll 1$, the trap charge is the dominant contribution to capacitive charging ($Q = CV$) per unit area ($A$) (i.e. $Q/A = \int (n_{TS} + n)dx = \int n_T(1 + \theta)dx \sim \int n_T dx$) due to applied bias. However, the current is only due to free carriers $n = \theta n_T$, i.e., a fraction of the trapped charge. Hence, the shallow-trap SCLC current ($I_{trap-S}$) is reduced by a factor $\theta$ from the trap-free current ($I_{trapfree}$) based on SCLC theory[16].

$$I_{trap-} = \theta I_{trapfree} ; \quad I_{trapfree} = \frac{9}{8}\frac{\mu\epsilon V^2}{L^3}A \quad \dots (2)$$

Further, the $IV$ characteristics "gently" transitions from shallow trap SCLC ($\alpha = 2$) via TFL to trap-free SCLC ($\alpha = 2$) as shown in Fig. 2(b). This is because traps near the injecting contact, remain above the "injecting" Fermi level and thus never get fully filled, as depicted in Fig. 2(a). For *deep traps SCLC*, unlike the valence band, trap occupation follows Fermi Dirac statistics instead of Boltzmann Distribution. Hence, deep trap energy ($E_{T-D}$) is such that $E_F - E_{T-D} < 3kT$ even at equilibrium and reduces further with bias. They are also fully filled at high bias, leading to prominent TFL, shown in Fig. 2(b). There are two implications for *deep trap SCLC* as indicated in Fig. 2(b). First, the $\alpha \approx 2$ regime (equivalent to shallow trap SCLC) is not observed for *deep* traps. Second, the $IV$ characteristics transitions directly from Ohmic

($\alpha = 1$), via a sharp TFL ($\alpha \gg 2$), to trap-free SCLC ($\alpha = 2$). Thus, *shallow* traps can qualitatively reproduce the HRS experimental *IV* characteristics where an increase in *shallow* trap density will reduce current levels progressively but preserve the $\alpha \approx 2$ dependence (unlike in *deep* traps). However, *shallow* traps do not exhibit sharp ($\alpha \gg 2$) TFL which is observed experimentally. Thus we propose an alternative to resolve the apparent inconsistency by using a single *shallow* trap model along with self-heating.

To *qualitatively* demonstrate the effect of self-heating on shallow trap-SCLC *IV* characteristics, we perform thermodynamic simulations (involving simultaneous heat and current transport modeling) which matches isothermal simulation at 298K at low voltage, shown in Fig. 2(c). However, beyond a critical power density, current shoots up akin to experimental behavior. The temperature ($T$) vs Voltage ($V$) plot in Fig. 2(c) shows that temperature shoot-up (above 298K) occurs simultaneously with current shoot-up ($\alpha \gg 2$) indicating correlation due to self-heating. To understand this current shoot-up we need to consider two aspects. First, the steady state temperature ($T$) depends upon power density ($p = JV$ i.e. power per unit area) by Eq. 3 under 1-D steady state conditions, derived from Fourier's law of heat conduction:

$$-\kappa \frac{d^2 T}{dx^2} = p/L \quad \ldots (3)$$

where $\kappa$ is thermal conductivity of the material. Upon integration with boundary conditions of $T = T_{ambient}$ at $x = 0$ and $x = L$, the peak temperature ($T_{max}$) occurs at $x = L/2$ such that

$$T_{max} - T_{ambient} = \frac{p}{\kappa} \frac{L}{8} \quad \ldots (4)$$

Thus, $T_{max}$ increases linearly with power density ($p$). As the current increases with bias (i.e. power density increase) an increase in temperature occurs, which is significant after a critical power level. Secondly, as the temperature rises, current increases as a response (observed experimentally and verified by simulations later). This further increases power density and hence temperature causing a positive feedback process. To eventually break the positive feedback, the current must become independent of the PCMO temperature. A metallic contact series resistance limited current will be effective for this purpose which will, in fact, reduce current at higher temperature in the contacts. Finally, compliance should limit the current.

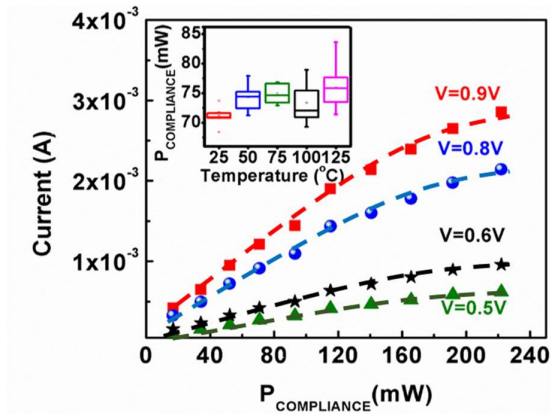

Fig 3: SCLC current vs. power at compliance indicating that SCLC current increases and then saturates as trap-free SCLC is achieved. Inset shows that $P_{COMPLIANCE}$ remains roughly equal with varying temperature.

To *quantitatively* model the electrical transport of the LRS, a further simplification is useful. We assume that LRS is trap-free based on the following rationale. During set (HRS →

LRS), the oxygen vacancies drift out of the PCMO device to reduce trap density. The trap density ($N_T$) reduction causes increase in SCLC current by a factor of $1/N_T$ (Eq. 2). The limit of current increase is trap-free SCLC beyond which no further current increase can occur. Oxygen vacancy velocity (drift) under electric field ($F$) estimated by Mott Gurney equation[25-26], given by

$$v = af e^{-\frac{E_m}{kT}} \sinh\left(\frac{F}{F_o}\right) \dots (5)$$

is strongly $T$-dependent (where critical field $F_o = 2kT/qa$, $a$ is hopping distance, $E_m$ is zero-field barrier height, $f$ is escape frequency). In Fig. 1(a), at ambient temperature ($T_{ambient} = 298K$), SCLC current level increases (resistance decreases) as $I_{COMPLIANCE}$ increases. The power at compliance ($P_{COMPLIANCE} = I_{COMPLIANCE} * V_{COMPLIANCE}$) produces $T_{max}$ (Eq. 4). Higher $T_{max}$ is related to more effective vacancy removal (Eq. 5) which should cause increase in current level until saturation sets in as all oxygen vacancies (i.e. traps) are removed and trap-free SCLC is achieved. This is observed experimentally in Fig. 3 where initially SCLC current level increases with increased $I_{COMPLIANCE}$ but then saturates to a trap-free LRS. This enables us to assume that the lowest LRS is trap-free at $T_{ambient} = 298K$. For higher $T_{ambient}$, $T_{max}$ increases with $T_o$ (Eq. 4) as power ($P_{COMPLIANCE}$) remains approximately same for same $I_{COMPLIANCE}$ (see Fig 3 inset). Hence oxygen vacancy motion is stronger at higher $T_{ambient}$. This implies that LRS is essentially trap-free at all $T_{ambient}$, at least above 298K.

To verify the model, we characterized our device at various $T_{ambient}$ ranging from 25°C-125°C. The experimental $IV$ characteristics of 10 sweeps at 25°C and 125°C are plotted in Fig. 4(c). The LRS shows the 4 regimes (discussed in Fig. 1(b)) in Fig. 4(c) with the series resistance regime exhibiting linearity for both temperatures (Fig. 4(c) inset) as in the intermediate temperatures (not shown).

We performed Sentaurus[TM] TCAD simulations where we use $\epsilon = 3000\epsilon_o$ which is consistent with a range of $\epsilon$ (i.e. $10^3$-$10^5$)[27]. Based on trap-free SCLC model, the experimental $IV$ characteristics at 298K is matched in regimes (i) and (ii) using $\mu = 3\ cm^2/Vs$. We assume a temperature dependence given by $\mu_T = \mu_{298K}(T/298)^{-\beta}$ where the exponent $\beta$ is assumed to be 2.2, based on values of other semiconductors like Si, GaAs and Ge[28]. To match the experimental Ohmic current temperature dependence, $\phi_B$ was varied to obtain a reasonable match for $\phi_B = 0.17eV$. Series resistance is extracted from experiments and inserted into the simulations. As PCMO has area-scalable current[2-3] contact area was estimated to be $1\ \mu m^2$ for probe tip with $14\mu m$ diameter (Fig. 4(b) inset).

To incorporate self-heating, the thermal resistance of the device i.e. (i) PCMO (ii) top electrode (i.e. W probe-tip) and (iii) bottom electrode (including substrate) was modeled by Finite Element Method(FEM) in MATLAB® in cylindrical coordinates, shown in Fig 4(b). An analytical model of the heat differential equation is used (i.e. $\kappa dT/dr = I_Q/A = IV/A$, where $I_Q$ is heat current in W, $\kappa$ is thermal conductivity in Wcm$^{-1}$K$^{-1}$, A is area of conduction). The thermal resistance is defined as $R_{TH} = dT/I_Q$. This is used to estimate the effective $R_{TH}$ of the top and bottom electrode to be $7\times10^3$ KW$^{-1}$ and $2.8\times10^4$ KW$^{-1}$ respectively which is incorporated into TCAD "thermodynamic" simulations that includes self-heating. The resistance of the PCMO layer ($4.2\times10^4$ KW$^{-1}$) is still comparable relative to electrode thermal resistances.

Based on this, the $IV$ characteristics in all regimes are simulated at different temperatures including regimes (ii) SCLC ($\alpha \approx 2$) (iii) thermal shoot-up ($\alpha \gg 2$) (iv) series resistance. Fig. 4(a) shows the experimental $IV$ characteristics exhibiting switching. In Fig. 4(c) we observe significant matching across all 4 regimes for 25°C and 125°C. Inset shows the linear plots of $I$ vs. $V$, which depicts significant matching of series resistance. Further, the simulated $IV$ at $T_{ambient}$ =25°C in Fig. 4(c) is identical to curve C in Fig. 2(c). The corresponding $T_{max}$ $vs. V$ at $T_{ambient}$ =25°C shows $T_{max}$ =438°C at 1.3V and 7.85mA by TCAD, which matches well with the FEM-based detailed T-profile calculation where $T_{max}$ =440°C (Fig. 4 (b)). Further, at $T_{ambient}$ =125°C, T-rise occurs at lower V than for $T_{ambient}$ =25°C.

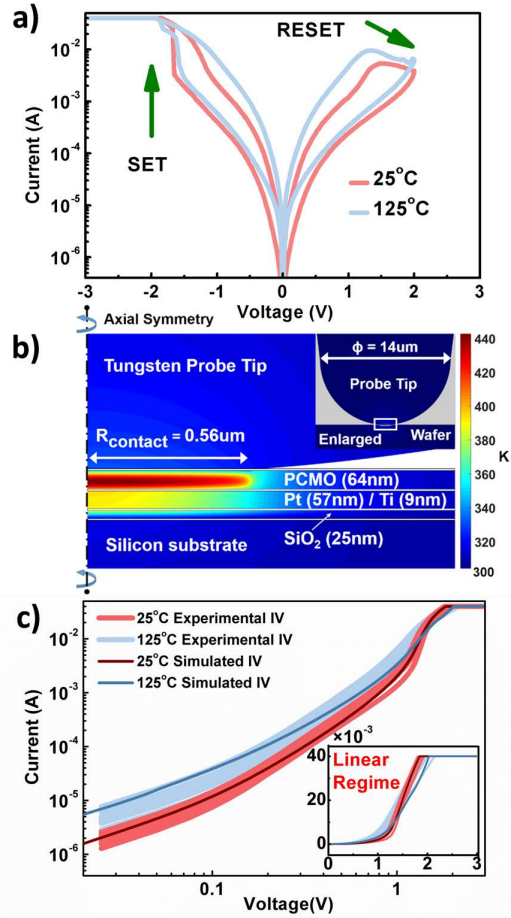

Fig 4: (a) $IV$ characteristics exhibiting switching at 25°C and 125°C (b) Cross-section of FEM thermal simulation of device, including Tungsten probe tip and substrate stack, showing the temperature distribution for V=1.3V, I=7.85 mA At 25°C, $T_{max} = 440°C$. The thermal conductivities in $Wm^{-1}K^{-1}$ are PCMO: 1.5[29], Pt: 71, W: 173, SiO$_2$: 1.3 and Si: 131[30]. Inset contains probe tip of diameter $14\mu m$ in contact with PCMO layer on wafer; (c) Log-Log and Linear (inset) experimental and simulated $IV$ plot for two different $T_{ambient}$ 25°C and 125°C across 10 voltage sweeps. Significant matching is observed in all 4 regimes.

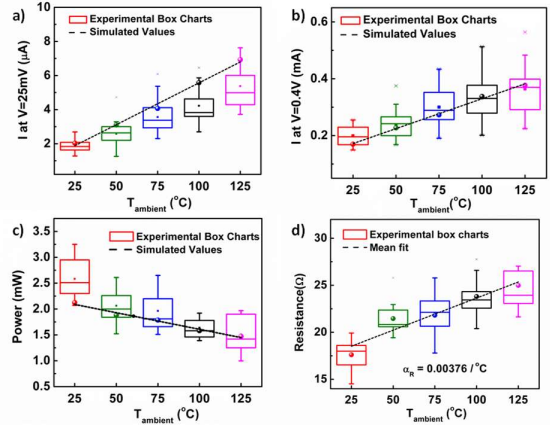

Fig 5: Comparison across various ambient temperatures ranging from 25°C-125°C between simulations and experimental $I$ at (a) V=25mV in Ohmic regime and (b) V=0.4V in SCLC regime exhibit significant matching. (c) Box charts of power ($P = I \times V$) at $V_{to}$ and simulated power values at corresponding temperatures displays significant correlation. (d) Box charts showing increase in resistance with increasing temperature from which temperature co-efficient of resistance is extracted from mean values.

In Fig. 5, experimental data and simulations are compared in each regime over temperatures ranging from 25°C-125°C. Fig. 5(a) and 5(b) respectively show the simulated dependence of Ohmic current (V=25mV) and trap-free SCLC (V=0.4V) is consistent with experimental observations at various $T_{ambient}$. This implies that a simple SCLC model can quantitatively predict experimental $IV$ characteristics in the Ohmic and SCLC regime including the temperature dependence.

For current shoot-up to commence, the internal device temperature $T_{max}$ shoot-up should initiate. As power $P \propto IV \propto V^3$ increases sharply with $V$, a critical power ($P_{crit}$) exists beyond which $T_{max} - T_{ambient} \gg 0$. We define the critical power $P_{crit} = V_{to}I_{to}$, where $V_{to}$ is take-off voltage and $I_{to}$ is the corresponding take-off current such that $\alpha$ increases from 2 by 10%. Fig. 5(c) shows linear relation between $P_{crit}$ vs. $T_{ambient}$ extracted from experimental data with a negative slope consistent with Eq. 3, which validates self-heating. In addition, simulations show excellent match with experiments.

Fig. 5(d) depicts the dependence of resistance extracted from the series resistance limited regime vs. $T_{ambient}$. We observe that resistance increases with temperature linearly. The temperature co-efficient of the series resistance ($\alpha_R$) is estimated to be $\alpha_R = 3.76 \times 10^{-3}/°C$ based on a temperature-dependent resistance model: $R_0(1 + \alpha_R(T - T_{ref}))$, where $T_{ref}$ is 20°C. This is similar to the range for metals e.g. Pt ($\alpha_R = 3.73 \times 10^{-3}/°C$) and W ($\alpha_R = 4.4 \times 10^{-3}/°C$)[31], indicating that the series resistance originates from metallic resistance.

Earlier[15], we have shown that negative and positive $IV$ characteristics in PCMO are symmetric in regime (i) and (ii). In regime (iii), negative LRS produced a current shoot-up while in positive LRS it initiates reset. The self-heating origin of current shoot-up in negative LRS is consistent with initiation of reset in positive LRS due to enhanced ionic transport by temperature change. Interestingly, more effective reset has been demonstrated experimentally in PCMO-RRAM by insertion of thermal insulators[2]. Further, thermally-assisted reset is proposed for non-filamentary RRAM[32]. Thus, the self-heating model to explain the DC characteristics provides a critical anchor-point for switching kinetics modeling in non-filamentary RRAM.

In this paper, we have shown that simple SCLC theory is not sufficient to capture the entire *IV* characteristics in PCMO-based RRAM devices, especially the sharp shoot-up in current which we show is erroneously attributed to trap-filled limit. We have proposed a model which uses simple *shallow* trap SCLC theory along with self-heating followed by series resistance dominated regime to quantitatively model the experimental *IV* characteristics in LRS over a range of ambient temperatures. Thus, this DC *IV* model provides a quantitative platform of further analysis of electron and ion dynamics in PCMO-based RRAM.